\documentclass[conference]{IEEEtran}
\IEEEoverridecommandlockouts
\usepackage{cite}
\usepackage{amsmath,amssymb,amsfonts}
\usepackage{algorithmic}
\usepackage{graphicx}
\usepackage{textcomp}
\usepackage{xcolor}
\usepackage{booktabs}
\usepackage{multirow}
\usepackage{array}
\usepackage{adjustbox}
\usepackage{stfloats}
\usepackage{comment}
\usepackage{diagbox}
\usepackage{tabularx}
\usepackage{hyperref}

\newcolumntype{Y}{>{\centering\arraybackslash}X}  
\newcolumntype{Z}{>{\raggedright\arraybackslash}X} 
\def\BibTeX{{\rm B\kern-.05em{\sc i\kern-.025em b}\kern-.08em
    T\kern-.1667em\lower.7ex\hbox{E}\kern-.125emX}}
\begin{document}

\title{Reducing Object Hallucination in Large Audio-\\Language Models via Audio-Aware Decoding}
\author{\IEEEauthorblockN{Tzu-wen Hsu}
\IEEEauthorblockA{ 
\textit{Purdue University}\\
West Lafayette, United States\\
gilberthsu123@gmail.com}
\and
\IEEEauthorblockN{Ke-Han Lu}
\IEEEauthorblockA{ 
\textit{National Taiwan University}\\
Taipei, Taiwan \\
d12942024@ntu.edu.tw}
\and
\IEEEauthorblockN{Cheng-Han Chiang}
\IEEEauthorblockA{
\textit{National Taiwan University}\\
Taipei, Taiwan \\
dcml0714@gmail.com}
\and
\IEEEauthorblockN{Hung-yi Lee}
\IEEEauthorblockA{
\textit{National Taiwan University}\\
Taipei, Taiwan\\
hungyilee@ntu.edu.tw}}


\maketitle

\begin{abstract}
Large Audio-Language Models (LALMs) can take audio and text as the inputs and answer questions about the audio.
While prior LALMs have shown strong performance on standard benchmarks, there has been alarming evidence that LALMs can hallucinate what is presented in the audio.
To mitigate the hallucination of LALMs, we introduce \textbf{Audio-Aware Decoding (AAD)}, a lightweight inference‐time strategy that uses contrastive decoding to compare the token prediction logits with and without the audio context. 
By contrastive decoding, AAD promotes the tokens whose probability increases when the audio is present.
We conduct our experiment on object hallucination datasets with three LALMs and show that AAD improves the F1 score by 0.046 to 0.428.
We also show that AAD can improve the accuracy on general audio QA datasets like Clotho-AQA by 5.4\% to 10.3\%.
We conduct thorough ablation studies to understand the effectiveness of each component in AAD.\footnote{\url{https://github.com/GillbertHsu/Audio-Aware-Decoding}}
\end{abstract}

\begin{IEEEkeywords}
Large audio-language models, Object hallucination, Contrastive decoding
\end{IEEEkeywords}

\section{Introduction}
Large Language Models (LLMs) have become foundational in natural language processing, demonstrating impressive capabilities in understanding and generating human-like text \cite{openai2024gpt4technicalreport, peng2023instructiontuninggpt4, yang2024qwen2technicalreport, grattafiori2024llama3herdmodels, geminiteam2025geminifamilyhighlycapable}. 
Recent advancements have expanded the use of LLMs to audio tasks by teaching LLMs to understand audio, resulting in Large Audio Language Models (LALMs)\cite{Yin_2024, caffagni-etal-2024-revolution, zhang-etal-2024-mm,rubenstein2023audiopalmlargelanguagemodel, 10389742, lu24c_interspeech, 10889444, arora2025landscapespokenlanguagemodels, ghosh2025audioflamingo2audiolanguage, arora2025landscape}. 
Although LALMs have shown promising performance on standard benchmarks that evaluate their audio comprehension ability~\cite{10448257,huang2025dynamicsuperb, sakshi2025mmau, yang2025sakuramultihopreasoninglarge, wang2024audiobench, chen2024voicebench, yang2025towards}, there has been evidence implying that they can hallucinate the contents in the audio context and do not truly understand the audio~\cite{kuan2024understanding, 10888384}.

Hallucination, where a model generates content that is not supported by the input or factual knowledge~\cite{huang2025survey}, is a critical safety issue for LLMs and large vision language models~\cite{zhang2023sirenssongaiocean, rawte2023surveyhallucinationlargefoundation, Huang_2025}.
Similarly, LALMs can also hallucinate objects that are not presented in the audio. 
\cite{kuan2024understanding} showed that an LALM might successfully describe an audio clip of a dog barking, but when asked if it hears a cat meowing, the model may hallucinate and respond "yes". 
\cite{kuan2024understanding} introduced a benchmark to evaluate LALM's performance on object hallucination tasks and provided a set of prompts shown to reduce hallucination. 
However, the performance of LALMs is highly sensitive to prompt design \cite{lu2025speechifeval}, making it difficult to craft prompts that generalize well across different LALMs and task scenarios. 
This limits the robustness of prompt-based approaches for mitigating hallucinations.

\begin{figure}[t!]
    \centering
    \includegraphics[width=0.5\textwidth]{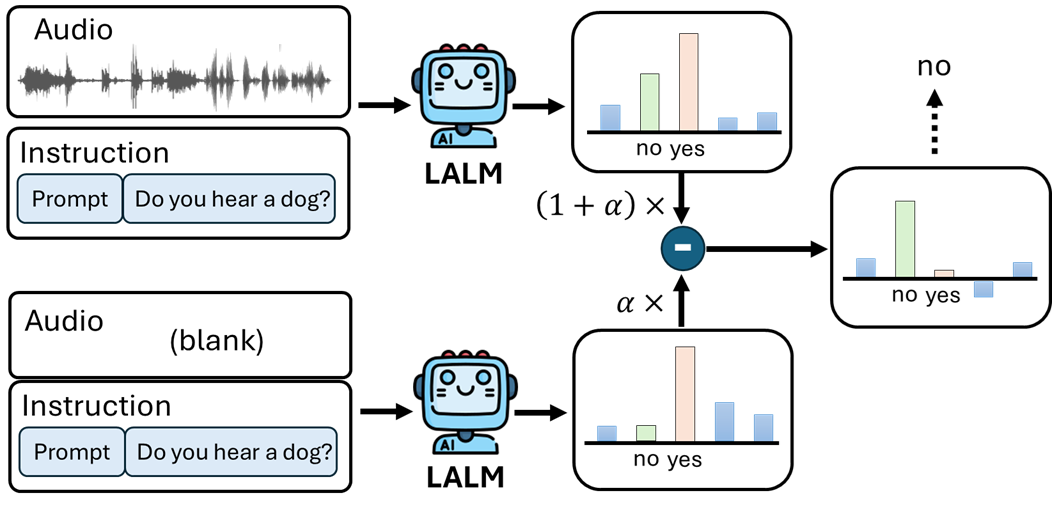}
    \caption{Illustration of our proposed Audio-Aware Decoding (AAD). 
    Since the probability of the “no” token increases significantly when audio is present, AAD up-weights this token, resulting in outputs that are more faithful to the audio context.}
    \label{fig:cad}
\end{figure}

Inspired by Context-Aware Decoding (CAD)~\cite{shi-etal-2024-trusting}, which was proposed to reduce hallucination of LLM under retrieval-augmented generation (RAG)~\cite{lewis2020retrieval,gao2023retrieval}, we propose an inference‐time method designed for LALM called \emph{Audio‐Aware Decoding (AAD)}.
AAD uses contrastive decoding~\cite{contrastivedecoding} to compare the token prediction logits when the audio is presented in the context and when there is no audio (silent audio) in the context.
By amplifying only those tokens whose probability increases when the actual audio is present, AAD effectively promotes the tokens that are grounded on the audio inputs.
An illustration of AAD is shown in Figure~\ref{fig:cad}. 

We conduct our experiment on the audio hallucination benchmark introduced by \cite{kuan2024understanding} using three LALMs, SALMONN-7B, SALMONN-13B \cite{tangsalmonn}, and Qwen2-audio-7B-Instruct \cite{chu2024qwen2audiotechnicalreport}, which differ in model sizes and architectures.
Our results show that AAD effectively suppresses object hallucinations and improves the F1 scores from 0.046 to 0.428 for different models and dataset subsets, providing a more robust and general approach than prompt engineering. 
We also conduct experiments on an audio QA dataset, Clotho-AQA \cite{9909680}. 
Although Clotho-AQA is not designed to evaluate object hallucination, we still found that AAD achieves improvements on this dataset ranging from 5.4\% to 10.3\%.
We include extensive ablation studies to understand each component of AAD. 
Our contributions are outlined as follows:
\begin{itemize}
    \item We introduce AAD, a contrastive decoding method for LALMs designed to emphasize the tokens whose probability is increased when the audio contexts are presented.
    To the best of our knowledge, this is the first application of contrastive decoding on LALMs.
    \item We show that AAD significantly reduces hallucination and even improves performance on general audio QA datasets.
    \item We conduct thorough ablation studies to understand each component in AAD.
\end{itemize}

\section{Method}
\label{sec: Method}

\subsection{Problem Formulation}
\label{sec:typestyle}
Our goal is to reduce object hallucination of LALM.
An LALM takes an audio clip and a corresponding question about that clip as input and produces a textual response. 
Object hallucination in an LALM occurs when the model is asked whether a certain object is present in the audio, and the model falsely claims that the object exists while it is not.
We introduce audio-aware decoding for LALMs, which adjusts the model’s token probabilities by contrasting its predictions with and without the audio context. 

\subsection{Audio-Aware Decoding (AAD)}
Before introducing AAD, we recap the standard decoding of LALMs.
In standard LALM decoding, the model generates each token autoregressively given the input question $\mathbf{x}$, the input audio $\mathcal{A}$, and the previously generated tokens up to time step $t$, denoted by $\mathbf{y}_{<t}$.
At decoding step $t$, LALM samples the next token based on the following distribution:
\begin{equation}
\mathbf{p}^{(t)}
\;=\;
\operatorname{softmax}\!\bigl[
\operatorname{logit}\bigl(y_t \mid \mathcal{A}, \mathbf{x}, \mathbf{y}_{<t}\bigr)\bigr]
\,.
\end{equation}

However, because the model can over‐rely on its model priors, e.g., what it has learned during training, it sometimes neglects the content in the input audio and hallucinates objects that do not actually appear in the audio. 
To address this issue, we introduce AAD, a contrastive decoding method that reduces the tendency to rely on the model's prior knowledge.

To resolve the model's over-reliance on its prior knowledge and neglect of the input audio, we use AAD to modify the token distribution during decoding.
Precisely, at each generation step $t$,  we feed two different inputs to the model separately: 
(1) \textbf{input with audio}: the LALM's input contains the audio $\mathcal{A}$, $\mathbf{x}$, and $\mathbf{y}_{<t}$, and (2) \textbf{input without audio}: the LALM contains a \textit{blank audio} $\mathcal{A}_{\text{blank}}$, $\mathbf{x}$, and $\mathbf{y}_{<t}$, where $\mathcal{A}_{\text{blank}}$ is an audio without any sound with the same length as $\mathcal{A}$.
This is done by creating a copy of the audio $\mathcal{A}$ with all zeros.

At each time step $t$, based on the two different inputs, the model produces two sets of logits:
\begin{itemize}
    \item $\text{logit}_{\text{with-audio}}^{(t)} = \text{logit}\big(y_t\mid\mathcal{A}, \mathbf{x},\mathbf{y}_{<t} \big)$: logits computed with the actual audio context $\mathcal{A}$.
    \item $\text{logit}_{\text{without-audio}}^{(t)} = \text{logit}\big(y_t\mid\mathcal{A}_{\text{blank}},\mathbf{x}, \mathbf{y}_{<t} \big)$: logits with the blank audio $\mathcal{A}_{\text{blank}}$.
\end{itemize}

We adjust the probability for each candidate token at step $t$ using the following contrastive decoding formula~\cite{contrastivedecoding, huang2025deltacontrastivedecoding, li2023contrastivedecodingopenendedtext}:

\begin{equation}
\label{eq: AAD}
\begin{split}
&\mathbf{p}^{(t)}_{AAD} \\
&= \text{softmax} \Bigl[(1 + \alpha)\, \text{logit}_{\text{with-audio}}^{(t)} \, -\alpha \, \text{logit}_{\text{without-audio}}^{(t)} \Bigr],
\end{split}
\end{equation}
where $\alpha$ is a hyperparameter that determines how much to emphasize the influence of the audio context versus the model's prior knowledge. 
A higher $\alpha$ places more weight on the audio context and downweights contributions from the model’s prior knowledge without hearing the audio.
At each time step, the token is sampled based on Equation~\ref{eq: AAD} and auto-regressively generates the output sequence.

The intuition behind AAD is straightforward: we examine how the likelihood of each candidate token changes when audio is introduced.
Tokens that become substantially more likely when the audio is presented are promoted, as they’re probably directly relevant to the given audio. 
By amplifying these context-sensitive signals, AAD steers the model toward more grounded and reliable outputs.

To help the LALM focus more on the audio content, we prepend a \textbf{\textit{prefix prompt}} before the question to tell the LALM to focus on the audio.
We use prompts provided by \cite{kuan2024understanding}, which is shown to reduce audio hallucination.
Unless otherwise specified, we use the following prompt: "\textit{Focus on the given audio and answer the following question}".

\section{Experiment setup}
\subsection{Evaluated LALMs}
We test AAD on three LALMs: SALMONN-7B, SALMONN-13B \cite{tangsalmonn}, Qwen2-Audio-7B-Instruct \cite{chu2024qwen2audiotechnicalreport}.
We select these models since they are instruction-tuned \cite{weifinetuned}, so they can follow instructions to answer the questions.
The three models also cover different sizes and model architectures, which can be used to validate whether AAD is general on diverse LALMs.

\subsection{Evaluated Datasets}

\paragraph{Audio Hallucination Dataset}
We use the object hallucination dataset from \cite{kuan2024understanding}.
Each instance in the dataset is a pair of audio and a yes/no question that asks if an object is in the audio snippet, such as “\textit{Is there a sound of [object] in the audio?}”, where \texttt{[object]} is a placeholder representing a candidate object. 
The \texttt{[object]} is sampled either from the set of ground-truth objects present in the audio or from a pool of objects not present in the audio.

Given an audio, different methods used to sample the absent \texttt{[object]} will create questions of different levels of difficulty.
\cite{kuan2024understanding} uses three sampling strategies to sample the absent objects for each audio clip: \textbf{\textit{random}}, \textbf{\textit{adversarial}}, and \textbf{\textit{popular sampling}}.

In all cases, absent objects are sampled from the set of all object labels present in the AudioCaps dataset \cite{kim-etal-2019-audiocaps}, excluding those that are present in the current audio clip.
In \textbf{random} sampling, absent objects are selected at random from all objects not present in the audio, providing a general measure of the model’s discrimination ability.
\textbf{Adversarial} sampling selects absent objects that most often co-occur with the true objects present in the audio. 
\textbf{Popular} sampling chooses the absent objects from the most frequent objects in the dataset. 
Each evaluation set is balanced, with an equal number of “yes” and “no” ground truth labels.
Refer to \cite{kuan2024understanding} for more details on this dataset.

\paragraph{\textbf{Clotho-AQA}}
In addition to the object hallucination benchmark, we evaluate all models on Clotho-AQA \cite{lipping2022clothoaqacrowdsourceddatasetaudio}, a standard audio question answering task. This benchmark measures the model’s broader audio reasoning capabilities and checks whether the application of CAD has any unintended effects on general audio comprehension and question answering performance. We select questions whose answers take the form of “yes” or “no” for convenience to assess the answer.

\renewcommand{\arraystretch}{1}
\begin{table*}[!ht]
    \centering
    \caption{Accuracy and F1 scores of various methods (Default, Prompt‐Only, AAD(\(\alpha\!=\!0.5\)), AAD(\(\alpha\!=\!1\))) across two datasets and an overall average.}
    \label{tab:equalwidth_no_tabularx_alpha1}
    
    \small
    \begin{adjustbox}{max width=\textwidth}
        \begin{tabular}{c cc cc cc cc cc}
            \toprule
            \multirow{3}{*}{\diagbox{\textbf{Method}}{\textbf{Dataset}}}
            & \multicolumn{6}{c}{\textbf{Audio Hallucination QA~\cite{kuan2024understanding}}}
            & \multicolumn{2}{c}{\multirow{2}{*}{\centering\textbf{Clotho-AQA}}}
            & \multicolumn{2}{c}{\multirow{2}{*}{\centering\textbf{Average}}}  \\
            \cmidrule(lr){2-7}
            & \multicolumn{2}{c}{\textbf{Random Sampling}}
            & \multicolumn{2}{c}{\textbf{Adversarial Sampling}}
            & \multicolumn{2}{c}{\textbf{Popular Sampling}}
            & 
            & 
            \\ 
            \cmidrule(lr){2-11}
            & Acc & F1
            & Acc & F1
            & Acc & F1
            & Acc & F1
            & Acc & F1 \\
            \midrule
            
            \multicolumn{11}{c}{\textit{\textbf{SALMONN‐7B}}} \\ [0.25em]
            \textit{Default    } 
                & 0.559 & 0.233 
                & 0.509 & 0.177 
                & 0.523 & 0.229 
                & 0.710 & 0.746 
                & 0.575 & 0.346 \\
            \textit{Prompt engineering    }
                & 0.689 & 0.564 
                & \textbf{0.511} & 0.324 
                & 0.574 & 0.464 
                & 0.743 & 0.772 
                & 0.629 & 0.531 \\
            \textit{AAD (Ours, \(\alpha=0.5\))}    
                & 0.733 & 0.661 
                & 0.505 & 0.416 
                & \textbf{0.575} & 0.540 
                & \textbf{0.764} & \textbf{0.790} 
                & \textbf{0.644} & 0.602 \\
            \textit{AAD (Ours, \(\alpha=1.0\))}    
                & \textbf{0.776} & \textbf{0.737}
                & 0.482 & \textbf{0.456} 
                & 0.548 & \textbf{0.562} 
                & 0.737 & 0.753 
                & 0.635 & \textbf{0.627} \\
            \midrule
            
            \multicolumn{11}{c}{\textit{\textbf{SALMONN‐13B}}} \\ [0.25em]
            \textit{Default    }     
                & 0.615 & 0.384 
                & \textbf{0.532} & 0.275 
                & \textbf{0.576} & 0.393 
                & 0.701 & 0.737 
                & 0.606 & 0.447 \\
            \textit{Prompt engineering   }
                & 0.678 & 0.533 
                & 0.521 & 0.256 
                & 0.555 & 0.372 
                & 0.758 & 0.804 
                & 0.628 & 0.491 \\
            \textit{AAD (Ours, \(\alpha=0.5\))}    
                & 0.720 & 0.621 
                & 0.519 & 0.323 
                & 0.568 & 0.439 
                & \textbf{0.773} & \textbf{0.815} 
                & 0.645 & 0.550 \\
            \textit{AAD (Ours, \(\alpha=1.0\))}    
                & \textbf{0.749} & \textbf{0.676} 
                & 0.514 & \textbf{0.356} 
                & 0.569 & \textbf{0.469} 
                & 0.773 & 0.801 
                & \textbf{0.651} & \textbf{0.575} \\
            \midrule
            
            \multicolumn{11}{c}{\textit{\textbf{Qwen2‐Audio‐7B Instruct}}} \\ [0.25em]
            \textit{Default    }
                & 0.568 & 0.302 
                & \textbf{0.487} & 0.247 
                & \textbf{0.505} & 0.298 
                & 0.711 & 0.778 
                & 0.568 & 0.406 \\
            \textit{Prompt engineering  }
                & 0.594 & 0.397 
                & 0.469 & 0.281 
                & 0.503 & 0.361 
                & 0.759 & 0.805 
                & 0.581 & 0.461 \\
            \textit{AAD (Ours, \(\alpha=0.5\))}        
                & 0.705 & 0.630
                & 0.458 & 0.347
                & 0.502 & 0.444 
                & \textbf{0.814} & \textbf{0.833} 
                & 0.620 & 0.564 \\
            \textit{AAD (Ours, \(\alpha=1.0\))}    
                & \textbf{0.762} & \textbf{0.737} 
                & 0.460 & \textbf{0.435} 
                & 0.504 & \textbf{0.506} 
                & 0.813 & 0.821 
                & \textbf{0.635} & \textbf{0.624} \\
            \bottomrule
        \end{tabular}
    \end{adjustbox}
\end{table*}

\subsection{Evaluation Metrics}

The datasets we use contain only yes/no questions.
For each question, we let the LALM freely generate the answer and use a rule-based parser to extract its final yes/no verdict.
We report \textbf{\textit{accuracy}} and \textbf{\textit{F1 score}} for both benchmarks. 

For the audio hallucination benchmark, our primary interest is in assessing the tendency of LALMs to incorrectly predict the presence of non-existent objects. 
Therefore, following \cite{kuan2024understanding}, we treat the answer "\textit{no}" as a "\textit{positive} instance" to compute recall and precision in the F1 score.
That is, if the model correctly says that an object is not in the audio when it is indeed not in the audio, this is considered as the "\textit{true positive}".
This setup allows us to specifically quantify the model’s ability to avoid hallucinating objects that are not present in the audio.
There are around 30,000 samples in this dataset.
Random guessing on this dataset will result in an accuracy of 0.500 and an F1 of 0.500.

For Clotho-AQA, evaluation follows the standard convention: the answer "\textit{yes}" is treated as a positive instance. 
There are 2490 samples for this dataset.
As this is also a class-balanced dataset, random guessing will yield an F1 and an accuracy of 0.500.

\subsection{Compared Methods}
We compare the following three methods:
(1) \textbf{Default}: the input only contains the question and the audio. 
(2) \textbf{Prompt engineering}: We prepend the prefix prompt, "\textit{Focus on the given audio and answer the following question},"  before the question.
This prompt was identified by \cite{kuan2024understanding} as yielding strong average performance on the original question.
We will sometimes refer to this baseline as "\textbf{prompt only}".
(3) \textbf{AAD}: the input contains the prefix prompt, the question, and the audio, and uses the contrastive decoding formula in Equation~\ref{eq: AAD}.
In our main experiments, we explore two values of $\alpha$: 0.5 and 1.0; we explore the sensitivity to $\alpha$ of AAD in section~\ref{subsection: alpha}.

\section{Results}
Table~\ref{tab:equalwidth_no_tabularx_alpha1} shows the experiment results across all sampling strategies in audio hallucination dataset and Clotho-AQA.
The tables report the accuracy (Acc) and F1 score (F1) across four configurations: Default, Prompt only, and AAD with different $\alpha$. 
In the following discussion on the hallucination dataset, we mainly focus on F1 rather than Acc.
Acc treats queries with 'yes' and 'no' as ground truth equally, while F1 places more weight on queries whose ground truth answers are "no".
Since this paper aims to reduce hallucination, we care more about whether LALMs can correctly respond with 'no' when the object is not present in the audio. Therefore, F1 is more important for the hallucination dataset.


\subsection{Random Sampling}
When absent objects are selected randomly, both Prompt-Engineering Only and AAD substantially improve over the Default baseline, with AAD consistently achieving the strongest results. For instance, on SALMONN-7B, the Default setting yields an F1 score of just 0.233. Prompt-Only significantly boosts this to 0.564 (+142\%), and AAD further elevates it to 0.737 (+216\%). Similar patterns occur with Qwen2-Audio-7B, where F1 improves from 0.302 (Default) to 0.397 (Prompt-Only), ultimately reaching 0.737 with AAD. Likewise, SALMONN-13B sees a notable increase from 0.384 (Default) to 0.533 (Prompt-Only) and finally to 0.676 with AAD. These large gains, especially AAD’s impressive 0.404 increase on SALMONN-7B, highlight how effectively AAD reduces hallucinations in scenarios where negative samples are randomly chosen.

\subsection{Adversarial Sampling}
In adversarial scenarios, Prompt-Only typically offers only minor benefits or, in some cases, even leads to performance drops. Despite this challenging context, AAD reliably recovers performance. Taking Qwen2-Audio-7B as an example, the Default F1 score is 0.247; Prompt-Only slightly improves this to 0.281, but AAD recovers substantially more, reaching 0.435 (+76\%). For SALMONN-7B, the Default setting's F1 of 0.177 improves notably to 0.324 with Prompt-Only, yet AAD surpasses this further, achieving 0.456 (+157\%). SALMONN-13B shows a similar trend: Default F1 is 0.275, Prompt-Only drops slightly to 0.256, but AAD recovers to 0.356 (+29.5\%). This clearly demonstrates that even when Prompt-Only struggles under adversarial conditions, AAD’s contrastive decoding mechanism effectively mitigates hallucinations.

\subsection{Popular Sampling}
When negative examples are selected based on popularity, the results closely mirror the adversarial scenario: Prompt-Only provides mixed results. It helps in some cases, but hurts performance in others. For example, Prompt-Only significantly boosts SALMONN-7B’s F1 from 0.229 (Default) to 0.464, yet it decreases SALMONN-13B’s F1 from 0.393 to 0.372. In contrast, AAD delivers consistent gains: SALMONN-7B's F1 soars from 0.229 to 0.562 (+145\%), Qwen2-Audio-7B climbs from 0.298 to 0.506 (+69.7\%). 
Thus, AAD maintains robust performance across varying model and demonstrates a greater robustness compared to using prompt engineering only.

\subsection{Clotho-AQA}
On the general audio-question answering task, Clotho-AQA, neither Prompt-Only nor AAD negatively impacts the models' ability to correctly answer the question. 
Instead, both approaches boost overall performance. Default F1 scores are already strong at 0.746 (SALMONN-7B), 0.737 (SALMONN-13B), and 0.778 (Qwen2-Audio-7B). Prompt-Only slightly enhances these to 0.772, 0.804, and 0.805 respectively, and AAD further lifts them to 0.790(+5.9\%), 0.815 (+10.6\%), and 0.833 (+7.1\%). This finding underscores that AAD's hallucination-reduction strategy does not compromise genuine question answering ability, but rather strengthens the model's ability to respond confidently when truly warranted by audio evidence.

\subsection{Overall Takeaway}
Across all tested conditions and models, AAD consistently surpasses both Default and Prompt-Engineering Only settings. 
In Random Sampling, AAD yields particularly impressive improvements, notably boosting SALMONN-7B’s F1 by 216\%. 
In more challenging conditions, such as Adversarial and Popular Sampling, where Prompt-Only's effectiveness wavers, AAD reliably restores and enhances performance by as much as 157\%. 
Crucially, on Clotho-AQA, AAD consistently improves or maintains accuracy, confirming that hallucination suppression via contrastive decoding does not diminish genuine audio question answering ability.

\section{Ablation study}

There are several components in AAD that are manipulable, including the hyperparameter $\alpha$ that controls the relative strength between the likelihood with and without the audio, and the \textit{\textbf{prefix prompt}} we use to make the model focus more on the audio.
In this section, we conduct ablation studies by varying the $\alpha$ and the prompts to understand their importance in AAD. 
We select Qwen2-Audio-7B-Instruct for our ablation analysis, as it is currently the most widely adopted LALMs.

\subsection{Ablation study on $\alpha$}
\label{subsection: alpha}
To examine the sensitivity of performance due to \(\alpha\), we evaluated the model in four values: 0.5, 1.0, 1.5, and 2.0 on object hallucination benchmark (random sampling) and Clotho-AQA. 
The results are shown in Table~\ref{tab:ablation-alpha}.
We report the accuracy and F1 as previously, and we additionally report the ratio of answering "\textit{yes}" to understand the model's behavior.

\begin{table}[t!]
\centering
\caption{Effect of varying \(\alpha\) in AAD on discriminative performance. Yes means the ratio that model outputs "yes"}
\label{tab:ablation-alpha}
\small
\setlength{\tabcolsep}{6pt}
\begin{tabular}{c|ccc|ccc}
\toprule
\multirow{2}{*}{$\alpha$}
  & \multicolumn{3}{c|}{\textbf{Random Sampling}}
  & \multicolumn{3}{c}{\textbf{Clotho-AQA}} \\
  & Acc & F1 & Yes(\%) 
  & Acc & F1 & Yes(\%) \\
\midrule
0   & 0.594 & 0.397 &  82.6
    & 0.759 & 0.805 &  67.3 \\
0.5 & 0.705 & 0.630 & 71.0
    & 0.814 & 0.833 & 55.5 \\
1.0 & 0.762 & 0.737 & 59.6
    & 0.813 & 0.821 & 48.6  \\
1.5 & 0.767 & 0.770 & 49.4  
    & 0.792 & 0.792 & 43.8 \\
2.0 & 0.738 & 0.764 & 39.0 
    & 0.751 & 0.733 & 37.1 \\
\bottomrule
\end{tabular}
\end{table}

The case when $\alpha$ is equivalent to no contrastive decoding, and we observe that the model is overly biased toward answering "yes," producing many false positives and thus lower F1 scores in object hallucination evaluation. 
Gradually increasing \(\alpha\) emphasizes more on the audio context and thus reduces this bias, as can be seen by the increase of the F1 score. 
The performance peaks around \(\alpha\) = 1.0, striking the best balance: the model significantly reduces incorrect "\textit{yes}" predictions (down to 59.6\% for Random Sampling and 48.6\% for Clotho-AQA) without losing its ability to correctly say '\textit{yes}' when the object the ground truth answer is '\textit{yes}'. 
Increasing \(\alpha\) beyond 1.0 yields diminishing returns; it further reduces the "\textit{yes}" rate but at the expense of overall F1 and accuracy. 
Consequently, for LALMs, the sweet spot appears around \(\alpha\) = 1.0, where the model best balances avoiding false “yes” outputs against still answering “yes” when appropriate.

When \(\alpha\) = 1.0 in AAD, the model’s inherent “yes” bias acquired during pretraining and fine-tuning gets effectively neutralized because the logits calculated with blank audio are subtracted equally from those calculated with real audio. In practice, this forces the LALM to only commit to a “yes” answer when the audio evidence is strong enough to overcome its prior inclination. As a result, the proportion of “yes” outputs shifts from around 90\% (under no AAD) or 70\% (under mild CAD) down to roughly 50\%. 
In general, we find that AAD with non-zero $\alpha$ can always improve over the default setting, while setting $\alpha=0.5$ or $1.0$ seems to be the best.

\subsection{Do We Need Prefix Prompts in AAD?}

In the design of AAD, we use a \textbf{\textit{prefix prompt}} ("\textit{Focus on the given audio and answer the following question}") to make the LALM pay attention to the audio.
Here, we explore whether this prefix prompt is necessary.
We conduct an experiment that removes this prefix prompt from the LALM input; in this case, the LALM input only contains the input question and the original or blanked audio when calculating the output token logits in Equation~\ref{eq: AAD}.

In Table~\ref{tab:cad-only}, we report the accuracy and F1 on the datasets when we do not use the prefix prompt.
We also report the performance relative to AAD with $\alpha =$ 0.5 in Table~\ref{tab:equalwidth_no_tabularx_alpha1}, which uses the prefix prompt.
We can see that removing the prompt generally reduces performance for most models under each benchmark, especially for SALMONN-7B. 
This behavior underscores the additive nature of AAD: the prompt steers the model’s attention toward the audio context, while AAD's contrastive decoding further amplifies tokens that truly depend on that context. 

\renewcommand{\arraystretch}{1}
\begin{table}[t!]
  \centering
  \caption{AAD Without Prompt Performance.
  The $\Delta$ is relative to AAD with $\alpha=$0.5 in Table~\ref{tab:equalwidth_no_tabularx_alpha1}.}
  \label{tab:cad-only}
  \small
  \begin{adjustbox}{max width=\columnwidth}
    \begin{tabular}{l|ccc}
      \toprule
      \textbf{Model} 
        & \textbf{Acc} & \textbf{F1} & \boldmath$\Delta$\textbf{Acc(\%)/$\Delta$F1(\%)} \\
      \midrule
      \multicolumn{4}{l}{\textbf{Random Sampling}} \\
      \midrule
      Qwen2‐Audio‐7B     & 0.656 & 0.555 & -4.9/-7.5 \\
      SALMONN‐7B         & 0.577 & 0.328 & -15.6/-33.3 \\
      SALMONN‐13B        & 0.706 & 0.604 & -1.4/-1.7 \\
      \midrule
      \multicolumn{4}{l}{\textbf{Adversarial Sampling}} \\
      \midrule
      Qwen2‐Audio‐7B     & 0.469 & 0.338 & +1.1/-0.9 \\
      SALMONN‐7B         & 0.502 & 0.276 & -0.3/-14 \\
      SALMONN‐13B        & 0.524 & 0.393 & +0.5/+7.0 \\
      \midrule
      \multicolumn{4}{l}{\textbf{Popular Sampling}} \\
      \midrule
      Qwen2‐Audio‐7B     & 0.484 & 0.374 & -1.8/-7.0 \\
      SALMONN‐7B         & 0.532 & 0.360 & -4.3/-18.0 \\
      SALMONN‐13B        & 0.586 & 0.516 & +1.8/+7.7 \\
      \bottomrule
    \end{tabular}
  \end{adjustbox}
\end{table}

\renewcommand{\arraystretch}{1}
\begin{table}[ht!]
  \centering
  \caption{Performance of Qwen2‐Audio‐7B under four configurations across three sampling strategies. Prompt-1 : "Focus on the given audio and answer the following question." Prompt-2: "Listen."}
  \label{tab:qwen2_audio_methods_alpha_comparison}
  \small
  \begin{adjustbox}{max width=\columnwidth}
    \begin{tabular}{l|cc|cc|cc}
      \toprule
      \multirow{2}{*}{\textbf{Method}}
        & \multicolumn{2}{c|}{\textbf{Random}}
        & \multicolumn{2}{c|}{\textbf{Adversarial}}
        & \multicolumn{2}{c}{\textbf{Popular}} \\
        & \textbf{Acc} & \textbf{F1}
        & \textbf{Acc} & \textbf{F1}
        & \textbf{Acc} & \textbf{F1} \\
      \midrule

      \textit{Default}     
        & 0.568 & 0.302 
        & 0.487 & 0.247 
        & 0.505 & 0.298 \\
        \midrule

      \multicolumn{7}{l}{\textbf{Prompt-1}: \textit{Focus on the given audio and answer ...} } \\
      \textit{Prompt engineering} 
        & 0.594 & 0.397 
        & 0.469 & 0.281 
        & 0.503 & 0.361 \\
      \textit{AAD ($\alpha=0.5$)}      
        & 0.705 & 0.630 
        & 0.458 & 0.347 
        & 0.502 & 0.444 \\
      \textit{AAD ($\alpha=1.0$)}      
        & 0.762 & 0.737 
        & 0.460 & 0.435 
        & 0.504 & 0.506 \\
      \midrule
      \multicolumn{7}{l}{\textbf{Prompt-2}: \textit{Listen.} }\\
      \textit{Prompt engineering} 
        & 0.538 & 0.202 
        & 0.503 & 0.220 
        & 0.512 & 0.255 \\
      \textit{AAD ($\alpha=0.5$)}      
        & 0.622 & 0.512 
        & 0.464 & 0.325 
        & 0.463 & 0.328 \\
      \textit{AAD ($\alpha=1.0$)}      
        & 0.651 & 0.658 
        & 0.457 & 0.446 
        & 0.454 & 0.444 \\
      \bottomrule
    \end{tabular}
  \end{adjustbox}
\end{table}

\subsection{Sensitivity to the Prefix Prompt}

Having seen that the introduction of the prefix prompt is important for AAD, we now ask if the performance of AAD is sensitive to the prefix prompt we use.
In this part, we use a different prefix prompt: "\textit{Listen}".
We want to understand if changing the prefix prompt can change the performance.
Recall that the original prefix prompt we used in Table~\ref{tab:equalwidth_no_tabularx_alpha1} is "\textit{Focus on the given audio and answer the following question}."
The prompt "\textit{Listen}" has been shown to be less helpful in a previous study~\cite{kuan2024understanding}.

As shown in Table~\ref{tab:qwen2_audio_methods_alpha_comparison}, using the prompt \textit{"Listen"} has a clear negative impact when it is used alone, as seen by the drop in F1 from 0.302 to 0.202 under Random sampling, from 0.247 to 0.220 under Adversarial, and from 0.298 to 0.255 under Popular. 
But even with this suboptimal prompt, AAD remains highly effective. 
With $\alpha = 0.5$, AAD boosts F1 to 0.512 (Random), 0.325 (Adversarial), and 0.328 (Popular), outperforming the weakened prompt. 
When $\alpha = 1.0$, performance improves even more, reaching F1 scores of 0.658, 0.446, and 0.444 across the three settings. 

Comparing the AAD using prompt 1 and prompt 2, we can observe some performance fluctuation, with AAD using prompt 2 underperforming AAD using prompt 1.
However, no matter which prefix prompt we use, the performance of AAD is always better than the default and the prompt engineering only baseline.
This is in stark contrast with the result that solely relies on prompt engineering, which can worsen the performance in some cases.
Our results show that while the performance of AAD can vary due to the selection of prompts, it consistently improves the performance, making it a more robust method compared with pure prompt engineering.

\section{Related work}

Our work focuses on reducing hallucination of LALMs using a contrastive decoding method \cite{contrastivedecoding}.
Most related to our work is context-aware decoding (CAD)~\cite{shi-etal-2024-trusting}, which is proposed to reduce hallucinations in retrieval-augmented generation \cite{gao2023retrieval}.
CAD reduces hallucination by using contrastive decoding to compare the token likelihood with and without the retrieved contexts.
While our method is inspired by CAD, we are the first to implement a contrastive strategy specifically for LALMs and show that it can effectively reduce object hallucination of LALMs.

Object hallucination in LALMs has received wide research attention recently.
Several works have addressed audio hallucination by fine-tuning models on curated or augmented datasets. 
For example, \cite{nishimura2024audiohallucinationslargeaudiovideo} study audio hallucinations in Video-LLAMA.
\cite{kuan2024understanding} identify object hallucination in LALMs and construct a dataset to evaluate hallucination in LALMs.
Recently, \cite{10888384} introduce another dataset to evaluate diverse aspects of hallucination of LALMs.


\section{Conclusion, Limitations, and Future Work}
In this paper, we introduce audio-aware decoding (AAD), a contrastive decoding method designed for reducing the hallucination of LALMs.
By comparing the output token logits when the input contains audio and does not contain audio, AAD promotes the token whose likelihood increases when the audio contexts are presented.
By extensive experiments with two datasets and three LALMs of various sizes and architectures, we show that AAD effectively reduces object hallucination and even improves general audio question answering ability, with an average boost of F1 score from 0.118 to 0.281.
We also conduct extensive ablation studies to understand the effect of each component in AAD to justify our current design.

While AAD is a promising method, there is an important limitation of this method:
For one question, we need to forward the through the LALM twice, one with the audio and one without it.
While this creates compute overhead compared to the default setting, the two forward passes in AAD can be done in parallel.
As a result, AAD should incur little to no latency compared with the default setting. We tested the inference speed on 3,000 data samples with and without AAD. When computation is not parallelized, inference time increases by nearly 60\%. However, when run in parallel, the increase is only around 19\%.
Based on the results in Table~\ref{tab:equalwidth_no_tabularx_alpha1}, trading off the inference compute overhead with the performance gain seems to be a reasonable choice.
Additionally, due to the nature of the datasets we use, we only evaluate object hallucination on yes/no questions.
It will be interesting to explore other types of object hallucination of LALMs by constructing more diverse benchmarks.
We leave this as future work.

Future work can also explore using different strategies to create the blank audio $\mathcal{A}_{\text{blank}}$, including replacing it with some non-zero constant audio or adding noise to the original audio.
While our focus is on using AAD to reduce hallucination, we also find that this can improve the performance on general audio question answering.
Future work can more systematically evaluate the performance of AAD on other audio-related question answering, including MMAU~\cite{sakshi2025mmau} or SAKURA~\cite{yang2025sakuramultihopreasoninglarge}.



\end{document}